\documentclass[journal=jpccck,manuscript=article]{achemso}
\usepackage[english]{babel}
\usepackage{graphicx}
\usepackage{epsfig}
\usepackage{color}
\usepackage{amsfonts,amssymb,amsmath}
\newcommand{\DIPC}[0]{{
Donostia International Physics Center (DIPC),
Paseo Manuel de Lardizabal 4, 20018 Donostia-San Sebasti\'an, Spain}}
\newcommand{\CFM}[0]{{
Centro de F\'{\i}sica de Materiales CFM/MPC (CSIC-UPV/EHU), Paseo Manuel de Lardizabal 5, 20018 Donostia-San Sebasti\'an, Spain}}
\newcommand{\ISM}[0]{{
 Universit\'e de Bordeaux, ISM, UMR 5255, F-33400 Talence, France}}
\newcommand{\ISMCNRS}[0]{{
 CNRS, ISM, UMR5255, F-33400 Talence, France}}

\author{O. Galparsoro}
 \affiliation{\DIPC}
 \alsoaffiliation{\ISM}
 \alsoaffiliation{\ISMCNRS}

\author{R. P\'etuya}
 \affiliation{\ISM}
 \alsoaffiliation{\ISMCNRS}
 \alsoaffiliation{\DIPC}

\author{J.\ I.\ Juaristi}
\affiliation{Departamento de F\'{\i}sica de Materiales, Facultad de
Qu\'{\i}micas (UPV/EHU), Apartado 1072, 20080 Donostia-San Sebasti\'an, Spain}
 \alsoaffiliation{\CFM}
 \alsoaffiliation{\DIPC}

\author{C. Crespos}
 \affiliation{\ISM}
 \alsoaffiliation{\ISMCNRS}

\author{M.\ Alducin}
 \affiliation{\CFM}
 \alsoaffiliation{\DIPC}

\author{P. Larr\'egaray}
\email{pascal.larregaray@u-bordeaux.fr / tel:+33540002961}
 \affiliation{\ISM}
 \alsoaffiliation{\ISMCNRS}

\title{ Energy Dissipation to Tungsten Surfaces Upon Eley-Rideal Recombination of N$_2$ and H$_2$}


\SectionNumbersOn 

\begin{document}
\maketitle
\begin{abstract}
Quasiclassical molecular dynamics simulations are performed to investigate
energy dissipation to the (100) and (110) tungsten surfaces upon Eley-Rideal (ER)
recombination of H$_2$ and N$_2$. Calculations are carried out within the single
adsorbate limit under normal incidence. A generalized Langevin surface oscillator
(GLO) scheme is used to simulate the coupling to phonons, whereas electron-hole
({\it e-h}) pair excitations are implemented using the local density friction
approximation (LDFA). Phonon excitations are found to reduce the ER reactivity for
N$_2$ recombination, but do not affect H abstraction. In contrast, the effect of {\it e-h}
pair excitations on the ER recombination cross section is small for N$_2$, but can be important
for H$_2$.  The analysis of the energy lost by the recombined species shows that
most of the energy is dissipated into phonon excitations in the
N$_2$ recombination
and into electronic excitations in the H$_2$ recombination. In all cases, the energy
dissipated into {\it e-h} pairs is taken away from the translational kinetic
energy of the formed molecules, whereas dissipation to phonons, only significant
for N$_2$, also affects vibration. Interestingly, the electron mediated energy
losses are found to be smaller in the case of N$_2$ when surface motion is allowed.

\end{abstract}


\begin{tocentry}
\begin{center}
\includegraphics[width=1.0\columnwidth]{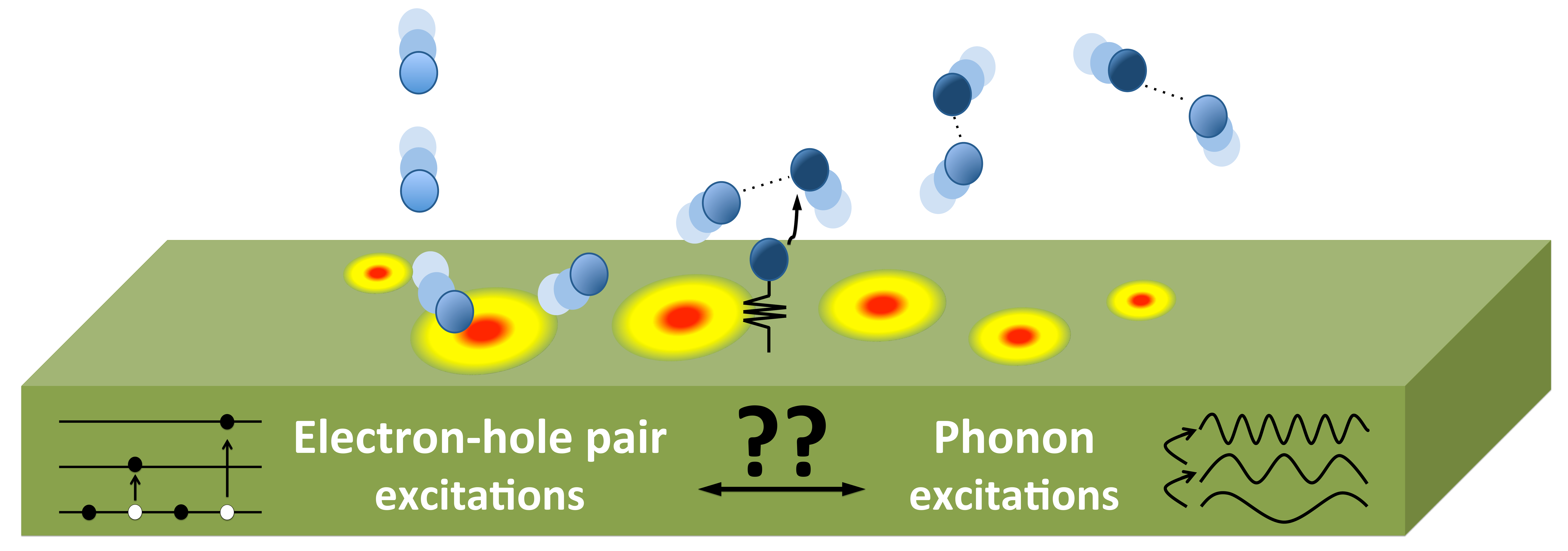}
\end{center}
\end{tocentry}
\newpage


\section{Introduction}\label{sec:sar}
The interaction of gas molecules and atoms with surfaces is of primary
importance in processes of practical interest such as
oxidation\cite{szekely2012gas}, corrosion\cite{pearson2014conservation}, hydrogen
storage\cite{Sakintuna20071121}, hetereogenous catalysis\cite{Honkala_2005,Rayment_1985,Somorjai_1994},
plasma wall interactions\cite{Federici_2002,Federici_2003}, chemistry
of atmospheric\cite{Greenberg_2002,Molina_1996} and interstellar
media\cite{Mathis_1993,Winnewisser_1993}. In the past years,
unprecedented accuracy has been achieved in the theoretical description of dynamics
of elementary processes at metal surfaces thanks to the development of \textit{ab
initio} calculations based on density functional theory (DFT) and the improvement in
computational capabilities. In that context, molecular dynamics simulations on
ground state multidimensional potential energy surfaces (PES) are widely used. Such
approaches, the validity of which breaks down when strong nonadiabatic couplings
take place upon gas-surface scattering,
prove to be very useful in the limit of weak
couplings.
In this context, previous works have shown that implementing
electron-hole ({\it e-h}) pair excitations in the dynamics was feasible.
Energy transfer to the metal electrons,
upon interaction/scattering of atoms and molecules, have been evidenced in a number
of experiments, \textit{e.g.}, in electric current measurements in Schottky
diodes\cite{Gergen21122001,Gergen2001123,Nienhaus1999-2,Nienhaus1999}
or metal-insulator-metal (MIM)
diodes\cite{Hasselbrink20091564,Hasselbrink2006192,Mildner2006133} during
exothermic reactions such as adsorption, abstraction, dissociation
and chemisorption; in the detection of particle emission such as
electrons (exoelectron emission)\cite{ROBERTSON1981} or photons (surface
chemiluminescence)\cite{Kasemo198023} in exothermic reactions; in measurements
of very short vibrational lifetimes of adsorbed molecules\cite{Krishna2006}; in
vibrational excitations measurements in the scattering of NO (HCl) molecules on
Ag(111), Cu(100) and Au(111)\cite{Rahinov2011}; and in vibrational relaxation
measurements in the scattering of vibrationally excited NO molecules on
Au(110)\cite{Rahinov2011}. Thus, the ensuing question to answer is whether such
energy transfers are relevant for each particular process under study.

One elementary process in gas-surface reactions that deserves special
attention is the recombination of gas-phase atoms with adsorbates.
Surface recombination proceeds through either
Langmuir-Hinshelwood, Hot-Atom (HA) or Eley-Rideal (ER) recombination. The
former, defined as the recombination between chemisorbed species, has been
found to play an important role in the recombination of heavy species such
as O and C \cite{Kori198332,Mullins1991,Wheeler1998,Wheeler1997}.
The recombination of light projectiles
(H,D)\cite{Kammler1997,Kim1999,Lykke1990,Rettner1992,Rettner21011994,Rettner1995,Guvenc2001,Lemoine2002,Martinazzo2004,Persson1995468,Quattrucci2005,Shalashilin1999},
on the contrary, is found to proceed in many cases via HA recombination, in
which the incoming gas-phase species experience few collisions with the surface
prior recombining with the adsorbate, at low collision energy and coverage. The
single-collision Eley-Rideal reaction cross sections are generally small,
with some exceptions \cite{Blanco-Rey2013,Ueta2011,Lemoine2002}. However,
it was recently highlighted that ER and HA mechanism compete in the
recombination of H on H-covered W(110) surface as collision energy or coverage
increases.\cite{Petuya2015}

ER abstraction has been theoretically scrutinized, mainly under the adiabatic approximation\cite{Persson1995,Lemoine2002,Guvenc2001,Bofanti2011,Blanco-Rey2013,Quintas2012,Petuya20142}, or only taking into account energy dissipation to the lattice phonons\cite{Shimokawa2000,Rutigliano2008,Quintas2013,Petuya2014,Kammler1999,Hansen1997L333,Guvenc2002,Cacciatore2004,Quintas-Sanchez2014}.
This process is of special interest as it is usually very exothermic and,
therefore, it provides highly excited
molecules\cite{Rettner1994,Rettner1992,Rettner1995} of potential
interest for negative ions productions \cite{Bechu2009}.
%
%
It was so far argued that {\it e-h} pair excitations should not play a major role
because of the ultrafast ER reaction times. Even though this seems reasonable, the
high amount of energy released in the reaction may substantially excite electrons
and, consequently, be transferred to the metal. For instance, experimental energy
distribution of the formed HCl molecules on Au(111)\cite{Rettner1994} are not
consistent with phonons excitations due to the large mass mismatch between H and Au.
Besides, a theoretical study of H$_2$ recombination on Cl covered gold surface has shown
that about half of the energy was released to the
adsorbates\cite{Quattrucci2005} but part of the
remaining energy loss was suggested to proceed via {\it e-h} pair excitations.

In the last years, different \textit{ab initio} theories have been developed
to deal with {\it e-h} pair excitations in molecular processes on metal
surfaces\cite{muino2013dynamics}. Among them, a good compromise between
accuracy of results and simplicity of implementation is offered by the
local density friction approximation (LDFA) \cite{Juaristi2008}, as shown in
ref~\citenum{Tremblay2010}.  Within this model, the energy dissipation in adsorption
\cite{Blanco-Rey2014, MartinGondre2012126}, scattering\cite{Martin-Gondre2012}
and dissociation\cite{Goikoetxea2009} processes on metal surfaces have been
analyzed. Moreover, the description of such processes using a full-dimensional
\textit{ab initio} PES and accounting for both dissipation channels has been
achieved by combining LDFA and the generalized Langevin oscillator model
(GLO) \cite{Tully1980} to incorporate energy exchange with the lattice phonons
\cite{Busnengo2004}.

Making use of this methodology, we here investigate the energy dissipation
due to both {\it e-h} pair and phonon excitations in four systems with very
different ER dynamics: (i) The abstraction of light H atoms on both W(100)
and W(110) crystallographic planes, where phonons excitations are expected
to be negligible\cite{Petuya2014,Petuya20142} and (ii) that of heavier N
atoms in the same surfaces, where the surface motion effects are significant
\cite{Quintas-Sanchez2014,Quintas2012,Quintas2013}.

The paper is structured as follows. Methodology and details of the calculations
are presented in section~\ref{sec:oin}.  In section~\ref{sec:ema}, the effects
of the energy dissipation on ER reactivity and on the final energy distribution
of the formed molecules are analyzed. Finally, we conclude in section~\ref{sec:ond}.

\section{Theoretical methods}\label{sec:oin}

The normal incidence scattering of atomic hydrogen and nitrogen off respectively
H- and N-preadsorbed W(100) and W(110) surfaces is investigated within the
zero coverage limit (single adsorbate), using quasiclassical
trajectories
(QCT), which takes into account the zero point energy (ZPE) of the
adsorbatei (see below).
Global ground-state PESs for two atoms interacting with an infinite and
periodic surface have been previously constructed from density functional theory
(DFT)\cite{Hohenberg1964,Kohn1965} calculations 
fitted by the flexible periodic London-Eyring-Polanyi-Sato
(FPLEPS)\cite{MartinGondre2009136,Martin-Gondre2010,MartinGondre2010136}
model. Details on the DFT calculations and the fitting procedure can be found
in refs~\citenum{Busnengo2008,Petuya2014} for H$+$H/W(100) and H$+$H/W(110), in
refs~\citenum{Martin-Gondre2010,MartinGondre2010136,Volpilhac2004129,Bocan2008}
for N$+$N/W(100), and in
refs~\citenum{Martin-Gondre2010,Bocan2008,Alducin2006PRL,Alducin2006JCP} for
N$+$N/W(110). The W(100) surface is known to undergo a structural phase transition
below 200~K. In this work, only the unreconstructed ($1\times1$) structure, which is
observed at temperatures higher than 200 K \cite{Ernst1992,Titmuss1996},
is considered. Therefore, temperature effects are examined above this value. In
order to rationalize nonadiabatic effects in the ER abstraction process, molecular
dynamics simulations have been performed within four different schemes:
\begin{enumerate}
\item The Born-Oppenheimer static surface approximation
(BOSS)\cite{Petuya20142,Quintas2013,Quintas2012}, in which neither energy
exchange with the surface phonons nor electronic excitations are
accounted for. 
\item The BOSS model including the effect of {\it
e-h} pair excitations within the LDFA
\cite{Juaristi2008}. In this model, electronic nonadiabaticity is introduced
through a dissipative force in the classical equations of motion for the
recombining atoms. This dissipative force is proportional to the atom velocity,
with a friction coefficient $\eta ({\bf r}_i )$, which depends on the vector
position ${\bf r}_i$ of the atom $i$. In the LDFA, the friction coefficient
applied to each recombining atom is that of the same atom moving in a homogeneous
free electron gas with electronic density equal to that of the bare surface
at the point ${\mathbf r}_{i}$ at which the atom $i$ is located. As explained
elsewhere \cite{Echenique1986,Echenique1981779}, the friction coefficients $\eta$
are evaluated to all orders in the atomic charge, in terms of the exact scattering
phase shifts at the Fermi level corresponding to the potential calculated within
DFT of a static atom embedded in the electron gas. In this way, the classical
equation of motion for each recombining atom reads,
\begin{equation}
 m_{i}\frac{d^{2}{\bf r}_{i}}{dt^{2}} = -{\bf \nabla}_{i}V({\bf
 r}_{i},{\bf r}_{j})-\eta({\bf r}_{i})\frac{d{\bf r}_{i}}{dt} \, ,
\end{equation}
where the first term in the right-hand side is the adiabatic force
obtained from the 6D PES $V({\bf r}_i, {\bf r}_j)$ and the second term
is the electronic friction force that accounts for the low-energy {\it
e-h} pair excitations. Note that the indexes $i$ and $j$
refer to the recombining atoms.
\item The Born-Oppenheimer moving surface approximation, which
introduces energy exchange with the metal phonons through a three-dimensional
surface oscillator connected to a thermal bath that accounts for energy
dissipation into the bulk (GLO
\cite{Adelman1979,Busnengo2004,Dohle1998,Polanyi1985,Tully1980})
%
\item The Born-Oppenheimer moving surface with electronic friction
(LDFA-GLO), which accounts for phonons and {\it e-h} pair excitations through GLO and LDFA\cite{Martin-Gondre2012}.
The combination of these two schemes allows us to analyze the competition between the two energy loss channels and their possible coupling.

\end{enumerate}

\begin{figure}
\includegraphics[width=0.75\columnwidth]{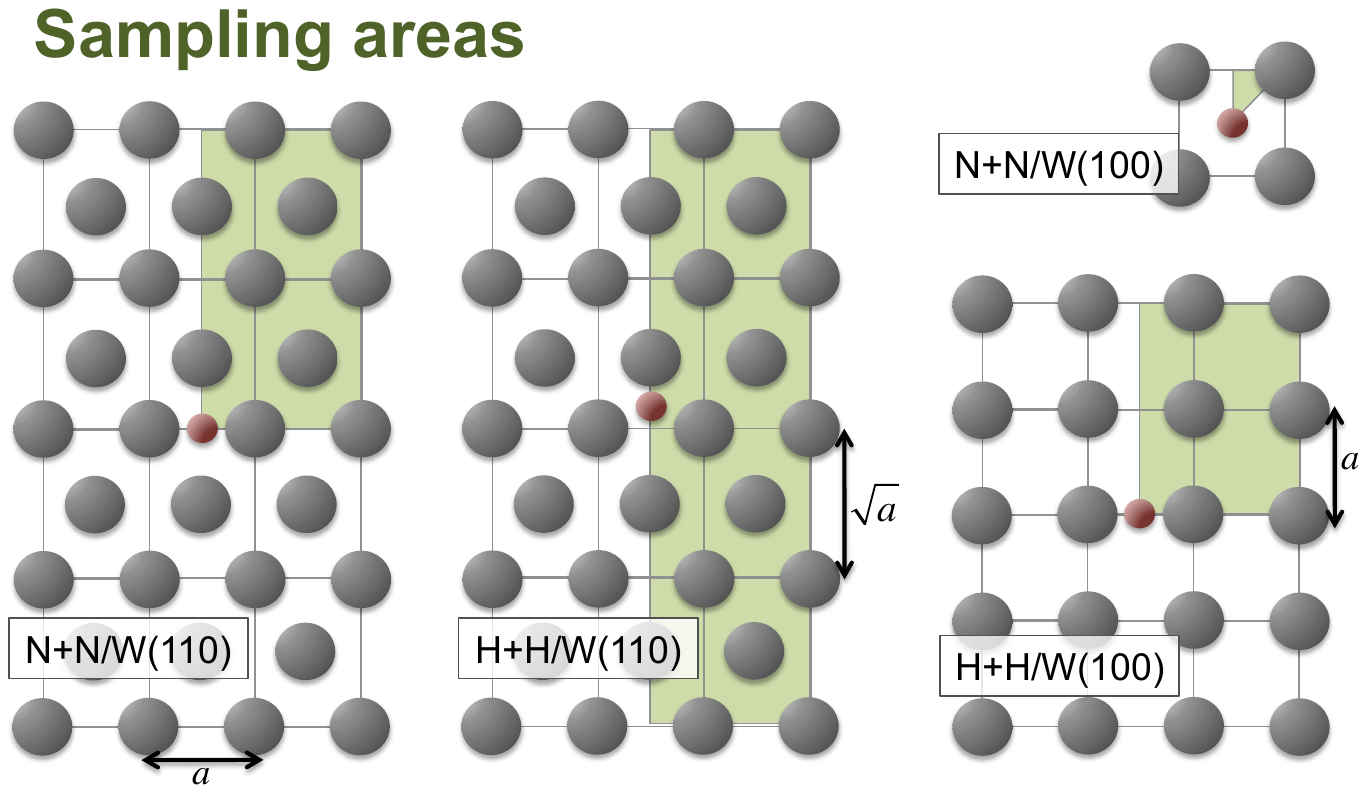}
\caption{Unit cell used for the dynamics simulations. The target atom is in red and the green area represents the sampling area of the ($X_{p}$,$Y_{p}$) initial position of the projectile.}
\label{fig:0}
\end{figure}

The initial conditions for the QCT simulations are described in the following. The adsorbed
atom (target) is initially at the most favorable adsorption site with velocity and
position consistent with the ZPE, which is calculated through a $x$,
$y$, and  $z$ 
mode decomposition (see table~\ref{tab:zpeadsit}). The ZPE values are in reasonable
agreement with experiments\cite{Balden1996,Balden1994,Barnes1978,Ho1980,Ho1978}.
In order to conserve the ZPE of the target before collision, the friction
force starts to act when target's energy exceeds the ZPE. The friction force is
then applied until the end of the trajectory. It was checked that this choice
makes negligible differences with respect to turning the friction force on only
when the target's energy exceeds the ZPE and turning it off when the target's
energy goes below the ZPE. The impinging H (N) atom, i.e., the projectile, starts
at $Z_p$ = 7.0~{\AA} (8.0~{\AA}), in the asymptotic region of the potential,
with normal incidence and initial collision energies E$_i$ that vary within
the range 0.25-5.0~eV. Taking advantage of the symmetry of each system, the
($X_p$,$Y_p$) initial coordinates of the projectiles are randomly sampled in
the green areas indicated in Figure~\ref{fig:0}. These areas differ from the
unit cells used in previous works \cite{Quintas2013,Quintas2012,Petuya20142}
because it was found that few ER recombination events occur for trajectories
starting outside of the unit cells used in previous works. Nevertheless, despite
some quantitative differences between the present and the previous
results, all the conclusions remain.

The ER cross section $\sigma_{ER}$ is defined by
\begin{equation}
 \sigma_{ER}=A\int\int_{D}P_{ER}(X_{p},Y_{p}) \, dX_{p}dY_{p} \, ,
\end{equation}
where the integration area $D$ is the sampling green area in Figure 1 and the factor
$A$ accounting for the total area per adsorbate is: $A = 4$ for H+H/W(100),
$A = 2$ for H+H/W(110), $A = 8$ for N+N/W(100) and $A = 4$ for N+N/W(110).
The two-dimensional opacity function $P_{ER}(X_p,Y_p)$ is the probability of
ER recombination for a given set of $X_p$ and $Y_p$, which defines the initial
position of the projectile.

For each collision energy, the number of computed trajectories was $368\,000$ for
H+H/W(100), $864\,000$ for H+H/W(110), $90\,000$ for N+N/W(100), and $900\,000$
for N+N/W(110). The possible exit channels of the simulations are defined in
detail elsewhere \cite{Quintas2013,Quintas2012}. Among them, the recombination
or abstraction processes studied here are considered to take place whenever
both atoms reach the initial
altitude of the projectile with a positive diatom center-of-mass momentum along the
surface normal ($z$-axis) and an interatomic distance $r<2.2$~{\AA} ($r<2.5$~{\AA})
in H$_2$ (N$_2$) recombination. The ER process occurs when the formed molecule moves
definitively towards the vacuum before the second rebound of the projectile.


\section{Results}\label{sec:ema}

\begin{figure}
\includegraphics[width=0.5\columnwidth]{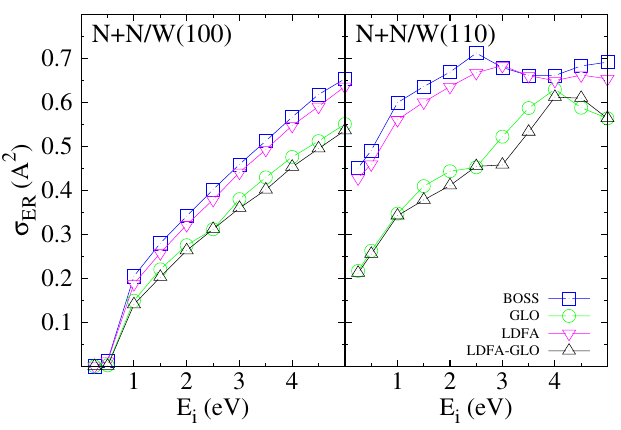}
\caption{Eley-Rideal recombination cross sections $\sigma_{ER}$ as a
function of the projectile's collision energy E$_i$ for the BOSS (blue
squares), LDFA (pink down-triangles), GLO (green circles) and LDFA-GLO
(black up-triangles) simulations. }
\label{fig:csN}
\end{figure}

\begin{figure}
\includegraphics[width=0.5\columnwidth]{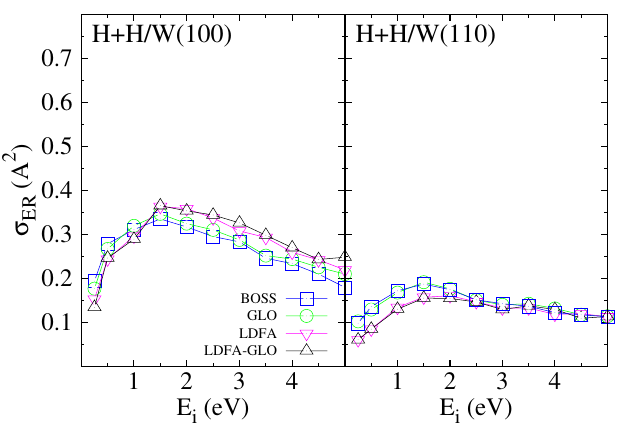}
\caption{Same as Figure~\ref{fig:csN} but for H$+$H/W(100) and H$+$H/W(110).}
\label{fig:csH}
\end{figure}

All the results shown in this section for the GLO and LDFA-GLO simulations
were obtained with a surface temperature of 300~K.  Simulations performed at
1500 K lead to similar results and are not shown. The computed ER cross sections
$\sigma_{ER}$ as a function of the projectile collision energy E$_i$ are displayed
in Figures~\ref{fig:csN} and \ref{fig:csH} for N$_2$ and H$_2$ recombination,
respectively.  Except for N$+$N/W(110), the qualitative behavior of $\sigma_{ER}$
with E$_i$ is almost unchanged by including the energy dissipation channels. The
largest quantitative differences respect to the BOSS results are found
for N$_2$
recombination. Figure~\ref{fig:csN} shows that surface motion is responsible for
reductions in the N$_2$ $\sigma_{ER}$ of 10-50\%, while electronic friction only
causes a marginal decrease of less than 7\%.  In the case of H$_2$ recombination
(Figure~\ref{fig:csH}), we observe just the opposite, i.e., the main changes respect
to the BOSS $\sigma_{ER}$ are due to {\it e-h} pair excitations (compare either LDFA
to BOSS or GLO to LDFA-GLO), while as also shown in ref~\citenum{Petuya20142} the
role of phonons can be disregarded. More precisely, including electronic friction
decreases the BOSS cross sections for the  H$+$H/W(100) in the range of 5-21\% at
low energies (E$_i <$1.0~eV), whereas at higher energies it increases in the range
of 8-22\%. For H$+$H/W(110), the effect of {\it e-h} pair excitations is only
observed at low energies (E$_i <1.5$~eV) with reductions in the range of 16-39\%.



As a general trend Figures~\ref{fig:csN} and \ref{fig:csH} reveal that
the ER reactivity decreases with electronic friction in the range
where $\sigma_{ER}$ increases with E$_i$ and increases when $\sigma_{ER}$ decreases
with E$_i$. Therefore, the effect of {\it e-h} pair excitations on $\sigma_{ER}$
is equivalent to shift the $\sigma_{ER}(\mathrm{E}_i)$ curves calculated within
the BOSS approximation towards smaller initial collision energies. This suggest
that the role of {\it e-h} pair excitation on $\sigma_{ER}$ is basically related
to the reduction of the collision energy.

The effect of the {\it e-h} pair excitations for N$_2$ dissociation on
W(110) \cite{Goikoetxea2009,Juaristi2008} and on W(100) \cite{Goikoetxea2009}, as well as for H$_2$ dissociation
on Cu(110)\cite{Juaristi2008} can be also bassically related to the reduction
of the collision energy. 
For the latter system, for which the dissociation is ruled by
a late activation barrier at short distances from the surface \cite{Salin2006},
a slight reduction of the (dissociative) sticking coefficient S$_0$ is predicted
when electronic friction is accounted for. This effect stems from the reduction
of collision energy, which prevents a non negligible portion of the molecules
to overcome the activation barriers. In contrast, for N$_2$ dissociation on
both tungsten surfaces, electronic friction causes in general a weak increase
of S$_0$. The reason is that N$_2$ dissociation is dominated in both surfaces
by early potential energy barriers lying in regions of low electronic densities.
Therefore, inclusion of electronic friction hardly affects the initial collision
energy before reaching the early barrier. However, once the molecules overcome
such early barriers and get close to the surface, {\it e-h} pair excitations
contribute to enhance the dynamic trapping towards the dissociation path by
slowing the molecules down.



In order to understand better the effect of each energy dissipation channel in
$\sigma_{ER}$,
the average dissipated energy $\left<\Delta
\mathrm{E}\right>$ as a function of E$_i$ obtained from the LDFA, GLO, and LDFA-GLO
simulations have been analyzed. In the latter case, the contribution coming solely from {\it e-h}
pair excitations $\Delta \mathrm{E}_{eh}$ is calculated for each trajectory as
\begin{equation}
\label{eq:elosseh}
\Delta \mathrm{E}_{eh}= \sum\limits_{i,n} \eta({\bf
r}_{i})\bigg(\frac{d{\bf r}_{i}}{dt}\bigg)^{2} \Delta t_n \, ,
\end{equation}
where the subscript $i$ refers to the projectile and target atoms and $\Delta
t_n$ is the time interval at the $n^{th}$ integration step. After averaging
$\Delta \mathrm{E}_{eh}$ over all ER trajectories, the average energy lost into
phonons $\left<\Delta \mathrm{E}_{\mathrm{ph}}\right>$ is obtained by substracting
$\left<\Delta \mathrm{E}_{eh}\right>$ to the total LDFA-GLO average energy loss $\left<\Delta
\mathrm{E}\right>$.

\begin{figure}
\includegraphics[width=0.5\columnwidth]{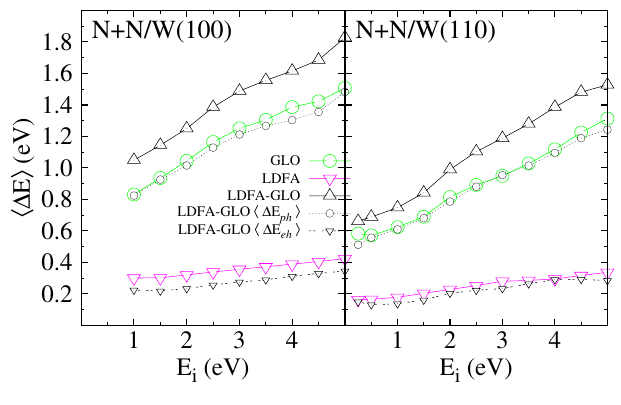}
\caption{Average energy loss $\left<\Delta\mathrm{E}\right>$ as a function of the
projectile's collision energy E$_i$ for the LDFA (pink down-triangles),
GLO (green circles) and LDFA-GLO (black up-triangles) simulations. For the
LDFA-GLO calculations, the average
energy loss into phonons $\left<\Delta\mathrm{E}_{ph}\right>$ (black
circles and dashed lines) and into
{\it e-h} pair excitations $\left<\Delta\mathrm{E}_{eh}\right>$ (black down-triangles 
and dashed lines) are also shown.}
\label{fig:elossN}
\end{figure}

\begin{figure}
\includegraphics[width=0.5\columnwidth]{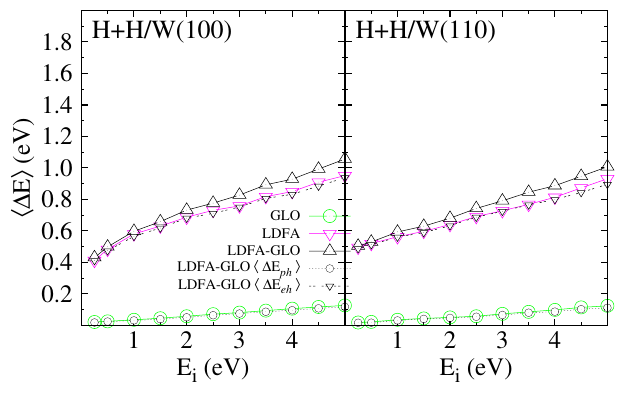}
\caption{Same as Figure~\ref{fig:elossH} but for H$+$H/W(100) and H$+$H/W(110).}
\label{fig:elossH}
\end{figure}

The results for N and H abstraction are displayed in Figures~\ref{fig:elossN}
and \ref{fig:elossH}, respectively. Comparing both figures, we observe that the
energy loss due to phonon excitations is one order of magnitude higher
for N$_2$
than for H$_2$ recombination. This is due to the smaller mass mismatch between
N and W than between H and W. In contrast, energy dissipation due to {\it e-h}
pair excitations is about three times larger for H$_2$ recombination
than for N$_2$
recombination, despite the friction coefficients at equal electron density are
significantly higher for N than for H \cite{Juaristi2008,Echenique1981779}. There
are two main factors contributing to this somewhat counterintuitive result. As
shown in Table~\ref{tab:rebz}, H atoms get closer to the surface and, therefore,
probe regions of higher electronic density than N atoms. In addition, for similar
collision energies the friction force and, hence, the electronic energy loss is
larger for H than for N due to the corresponding higher velocity of the former.


Focusing on the H$_2$ recombination, Figure~\ref{fig:elossH} shows that
almost all the energy loss is due to {\it e-h} pair excitations. Identical
conclusion was recently reported for the relaxation of the ``hot H
atoms'' formed
from H$_2$ dissociation on Pd(100) \cite{Blanco-Rey2014}.  In the case of
N$_2$ recombination, however, phonons is the predominant energy loss channel
as evidenced in Figure~\ref{fig:elossN}. Interestingly, there is an apparent
coupling between the two dissipation mechanisms, which is not observed
for H$_2$
recombination. Comparing the LDFA $\left<\Delta\mathrm{E}\right>$, on the one
side, and the GLO $\left<\Delta\mathrm{E}\right>$, on the other side, to the
LDFA-GLO results, we observe that the energy dissipated into either {\it e-h}
pairs ($\left<\Delta \mathrm{E}_{eh}\right>$) and into phonons ($\left<\Delta\mathrm{E}_{ph}\right>$) is
smaller in the LDFA-GLO simulations that account for both dissipation channels.
In relative terms, the largest effect is observed in the electronic dissipation channel, for
which there is a reduction of $\sim 0.1$~eV on N$+$N/W(100) and $\sim 40$~meV on
N$+$N/W(100) respect to the average energy loss obtained with the LDFA simulations. 
The existence of such a coupling between the the two energy dissipation
mechanisms contrasts with what it is observed not only for H$_2$ recombination,
but, importantly, for other processes involving N projectiles. For instance,
the competition between electron and phonon excitations in the scattering of
nitrogen atoms and molecules off tungsten and silver surfaces was analyzed in
ref~\citenum{Martin-Gondre2012}. As in the present work, it was found that
dissipation to surface vibrations was the predominant dissipation channel,
but at variance with our findings, the contribution of phonon and {\it e-h}
pair excitations to the total energy loss were shown to be basically additive.

\begin{figure}
\includegraphics[width=0.5\columnwidth]{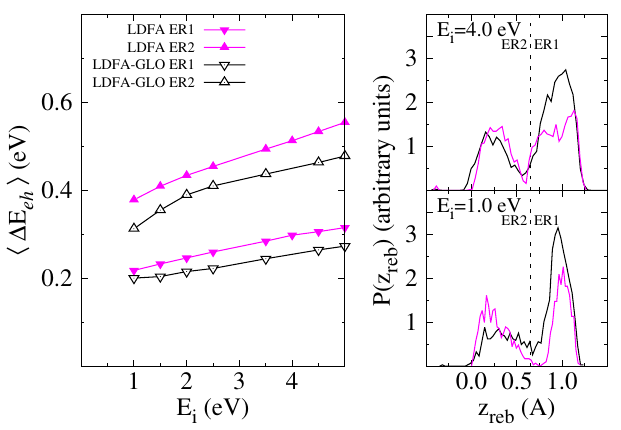}
\caption{Right panel: Average energy loss due to {\it e-h} pairs $\left<\Delta\mathrm{E}_{eh}\right>$ of molecules formed
via ER1 (up-triangles) and ER2 (down-triangles) as a function of the
initial collision energy of the projectile E$_i$ for the LDFA (pink) and
LDFA-GLO (black) simulations. Left panel: Distribution of the projectile
rebound altitude obtained from LDFA (pink) and LDFA-GLO (black) calculations.}
\label{fig:zrebN}
\end{figure}

In order to understand the reasons causing the coupling between the two dissipation
channels, we analyze in more detail the differences between the LDFA and LDFA-GLO
simulations for the N+N/W(100) system, for which the consequences of the coupling
between the energy loss channels are clearly more pronounced. Quintas {\it et
al.}~\cite{Quintas2012} identified two distinct ER abstraction mechanisms for
N$_2$ recombination on W(100), namely, one denoted ER1, which is characterized by
a projectile rebound altitude $Z_{\mathrm{reb}}$ higher than 0.65~{\AA},
and another denoted ER2, for which $Z_{\mathrm{reb}} <0.65$~{\AA}. Important to
us, the authors found by comparing the BOSS and the GLO results that the ratio
between ER1 and ER2 changes when surface motion is included. Here we find something
similar when comparing the LDFA and LDFA-GLO results. The distributions of the
projectile rebound altitudes for ER reactive trajectories are displayed in the
right panels of Figure~\ref{fig:zrebN} at two representative collision energies
E$_i= 1.0$~eV and 4.0~eV. The weight of ER1 increases from 49\% in LDFA to 79\% in
LDFA-GLO for E$_i$=1.0~eV (from 57\% to 71\% for E$_i$=4.0~eV). This modification
will certainly contribute to a reduction in the electronic energy loss, since the
molecules formed via ER1 probe surface regions of smaller electronic density than
molecules formed via ER2. Nevertheless, this is not the only ingredient that causes
the differences between the LDFA and the LDFA-GLO electronic energy losses. The
left panel of Figure~\ref{fig:zrebN} shows the average energy loss due to {\it e-h}
pair excitations for each mechanism within the LDFA and LDFA-GLO simulations. This
figure highlights that there is already a decrease in the electronic energy loss of
approx. $40$~meV in each mechanism when surface motion is included. Interestingly,
the reduction is similar to the one found for the N$+$N/W(110) system.  In this
case, we find that the $Z_{\mathrm{reb}}$-distribution of the ER recombinations
remains unchanged when including surface motion (not shown). Therefore, the
electronic energy loss decreases more on W(100) than on W(110) because in addition
there is a change in the $Z_{\mathrm{reb}}$-distributions of the former. Still,
the question that remains is why there is a systematic reduction in the electronic
energy loss when energy exchange with the lattice is allowed.

\begin{figure}
\includegraphics[width=0.5\columnwidth]{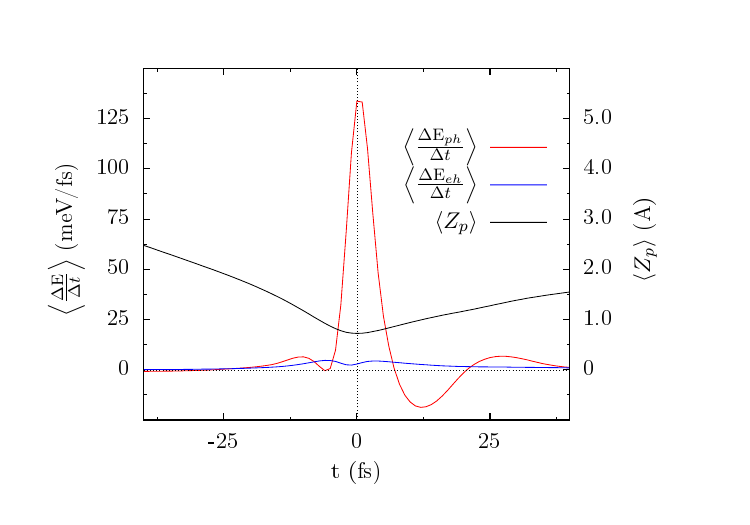}
\caption{Average energy loss rate of 300 ER trajectories due to phonons
($\left<\frac {\Delta \mathrm{E}_{ph}}{\Delta t}\right>$, in
red) and to {\it e-h} pairs ($\left<\frac {\Delta
\mathrm{E}_{eh}}{\Delta t}\right>$, in blue).
The right $y$-axis indicates the values of the average $Z$-coordinate 
of the projectile $\left<Z_p\right>$ (in black).}
\label{fig:Elossvst}
\end{figure}

With this purpose, we have analyzed the time evolution of each energy loss process
along the ER trajectories. Figure~\ref{fig:Elossvst} displays the average energy
loss rate to phonons $\left<\frac {\Delta \mathrm{E}_{ph}}{\Delta t}\right>$ and to metal
electrons $\left<\frac {\Delta \mathrm{E}_{eh}}{\Delta t}\right>$ as a function of time
for the N$+$N/W(100) system and E$_i=1.5$~eV. For each ER trajectory, the energy
loss rates are calculated by evaluating at each integration step $\Delta t$ the
contribution of each energy loss channel following the scheme explained above
(see eq~(\ref{eq:elosseh}) and text). The results of Figure~\ref{fig:Elossvst}
are averaged over 300 trajectories, after setting in each case the time origin
($t=0$) at the instant of the projectile's rebound.
As shown in the figure, most of the energy dissipated into the surface lattice
occurs at the classical turning point (see the large symmetric peak centered at
$t\approx 0$~fs that amounts about 0.76~eV). Afterwards, the forming molecule gains
and loses energy, but the energy exchange
in these cases is considerably smaller. For instance, the energy gain centered
at $t\approx 10$~fs is about 0.15~eV. At first sight, the electronic energy loss
rate, which vanishes at the classical turning point with the $z$-component of the
projectile's velocity, is rather symmetric around this point. The latter suggests
that the surface electron density and the N atoms velocities are rather similar
along the incoming ($t<0$) and outgoing ($t>0$) parts of the trajectory.

\begin{figure}
\includegraphics[width=0.5\columnwidth]{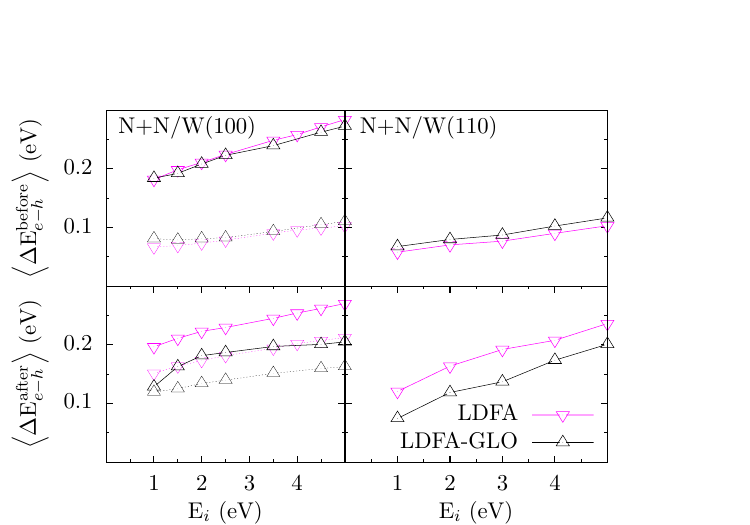}
\caption{Average energy loss due to {\it e-h} pair excitations before
($\left<\Delta\mathrm{E}_{eh}^{\mathrm {before}}\right>$, top panels) and after ($\left<\Delta\mathrm{E}_{eh}^{\mathrm {after}}\right>$, bottom
panels) the first collision with the surface. Left panels: results for the ER1
(solid lines) and ER2 (dashed lines) mechanisms identified in N$+$N/W(100). Right
panels: results for all ER trajectories in the case of N$+$N/W(110). Results from
the LDFA-GLO (LDFA) calculations are plotted in black (pink).}
\label{fig:Elossvsei-AB}
\end{figure}

The analysis of Figure~\ref{fig:Elossvst} highlights that, when phonons are
accounted for, the projectile loses an important part of its kinetic energy upon
the first collision with the surface. Consequently, in comparison with the static
surface LDFA calculation, the electronic friction force and, correspondingly,
the energy loss are expected to decrease for the remaining (outgoing) part of the
trajectory. In order to confirm the latter, we show in Figure~\ref{fig:Elossvsei-AB}
the energy loss into {\it e-h} pairs of the ER trajectories before
($\left<\Delta\mathrm{E}_{eh}^{\mathrm{before}}\right>$) and after
($\left<\Delta\mathrm{E}_{eh}^{\mathrm{after}}\right>$)
the first impact with the surface for N$+$N/W(100) and for N$+$N/W(110). In
all cases, the differences between the LDFA and LDFA-GLO calculations only are
significant after the collision event.  Obviously, since both dissipation channels
depend and act on the kinetic energy of the moving species, the inclusion of one
affects the other. Nevertheless, we have shown here that this
coupling is relatively small even when both dissipation channels are of the same
order of magnitude.  Otherwise, its effect will be imperceptible as found in
previous works \cite{Blanco-Rey2014,Martin-Gondre2012}, as well as in the present
work for H$_2$ recombination.



\begin{figure}
\includegraphics[width=0.5\columnwidth]{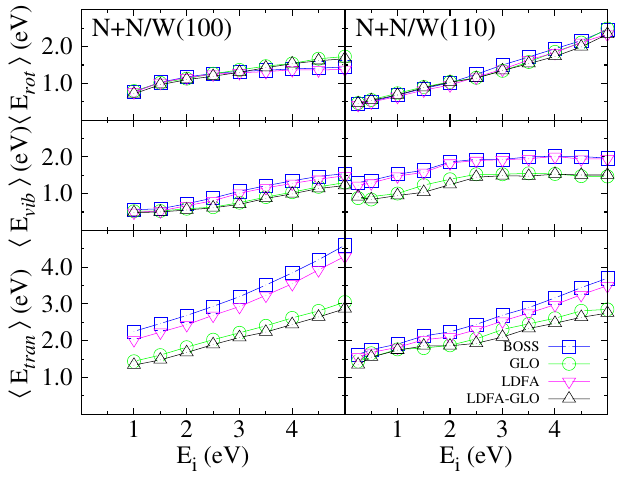}
\caption{Final average translational $\left<\mathrm{E}_{trans}\right>$, vibrational $\left<\mathrm{E}_{vib}\right>$, 
and rotational energies  $\left<\mathrm{E}_{rot}\right>$ of the formed molecules as a function of the
initial collision energy E$_i$: BOSS (blue squares), LDFA (pink down
triangles), GLO (green circles), and LDFA-GLO (black up triangles).}
\label{fig:ieN}
\end{figure}

\begin{figure}
\includegraphics[width=0.5\columnwidth]{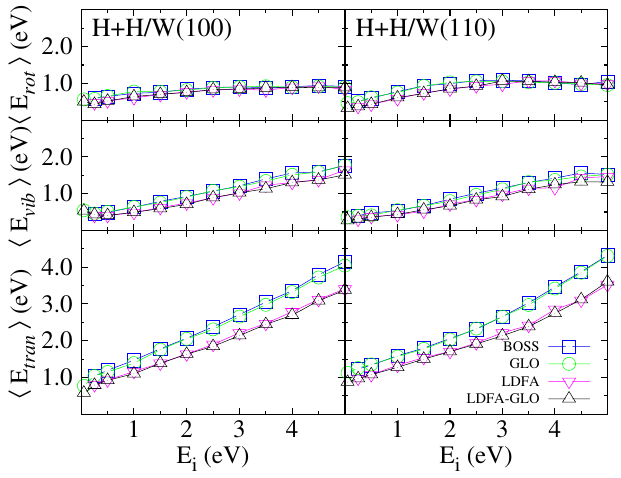}
\caption{Same as Figure~\ref{fig:ieN} but for H$+$H/W(100) and H$+$H/W(110).}
\label{fig:ieH}
\end{figure}

Finally, we analyze the changes that the two energy dissipation channels
may have in
the internal energy of the formed molecules. The final average translational,
vibrational, and rotational energies of the ER-formed N$_2$ and H$_2$ are plotted
in Figures~\ref{fig:ieN} and \ref{fig:ieH}, respectively, as a function of the
initial collision energy E$_i$. As shown in Figure~\ref{fig:ieN}, the largest
effect on N$_2$ ER recombination are due to phonon excitations, which affect both the
vibrational and the translational energy, as already discussed in
ref~\citenum{Quintas2013}. Our new simulations that include energy dissipation into
{\it e-h} pair excitations show that this mechanisms causes a decrease of the
translational energy only. However, such a decrease is small in comparison to
the reductions caused by phonon excitations. Regarding H$_2$
recombination, Figure~\ref{fig:ieH} shows that the effect of
energy dissipation into the metal electrons in the vibrational and rotational
energies is very minor, but noticeable on the translational energy, which
is reduced by 0.25-1.0~eV. In agreement with the results discussed so far,
the effect of surface phonons is negligible in both the translational and the
internal energy of the formed H$_2$.

\section{Conclusions}\label{sec:ond}

We have performed quasiclassical molecular dynamics simulations
allowing us to disentangle the influence of electron-hole pair and
phonon excitations on the Eley-Rideal recombination of H$_2$ and N$_2$ on the
(100) and (110) crystallographic planes of tungsten. Calculations are carried
out within the single adsorbate limit under normal incidence condition in the
0.25-5.0~eV energy range. Energy transfer due to phonon excitations is described
within the generalized Langevin oscillator scheme \cite{Tully1980,Busnengo2004}
and electron-hole pair excitations are modeled within the local density friction
approximation \cite{Juaristi2008}.

We confirm that phonon excitations reduce reactivity in the case of
N$_2$
recombination in the range of 10-50\% depending on the incidence energy and
the crystal face. However, phonon excitations do not affect H abstraction
due to the large mismatch between the mass of the projectile and that of the
tungsten atoms. Regarding electron-hole pair excitations the opposite behavior is
observed. Whereas they have a very minor effect on the Eley-Rideal recombination of
N$_2$, they can produce variations of the cross section for H$_2$ recombination of
up to a 36\%. The effect of including electronic excitations in the dynamics can
be rationalized as a reduction of the effective collision energy. As a result, in
the regions where the Eley-Rideal cross sections increase (decrease) with energy,
electronic excitations reduce (enhance) the recombination probability.

We have also evaluated the energy exchanged between the molecule and the metal
separating the contributions of each of the dissipation channels.  We find that
whereas energy loss due to electron-hole pair excitations is about three times
larger for H$_2$ recombination than for N$_2$ recombination, energy loss due to
phonons is an order of magnitude larger for the latter than for the former. Although
phonons are the main energy loss channel for N$_2$ formation, we observe that
electronic excitations are not negligible in this case. However, in the case of
H one can safely neglect any effect related to phonon excitations. Finally, we
have analyzed how the energy losses are distributed among the different degrees
of freedom of the molecules. We observe that electron-hole pair excitations mostly
reduce the translational energy of the molecules, whereas phonon excitations (only
significant in the case of N$_2$ recombinations) also affect the vibrational energy.

All in all,
 the description of the Eley-Rideal process is here refined
by including dissipation channels in the dynamics.  We have demonstrated
that in the case of H$_2$ recombination, due to its light mass, it is enough
to incorporate electron-hole pair excitations and that surface movement can be
neglected. On the contrary, for N$_2$ recombination, phonon excitations is the
dominant mechanism, though a noticeable effect of the electronic excitations is
also obtained.

\begin{acknowledgement}
O.G., J.I.J, and M.A. acknowledge financial support by the Basque
Departamento de Educaci\'on, Universidades e Investigaci\'on, the University
of the Basque Country UPV/EHU (Grant No IT-756-13) and the Spanish
Ministerio de Econom\'ia y Competitividad (Grant No.
FIS2013-48286-C2-2-P). O.G., M.A., and P.L. acknowledge the IDEX
Bordeaux ( ANR-10-IDEX-03-02 ) and Euskampus for fundings.
Computational resources were provided by the DIPC computing center and
the Mésocentre de Calcul IIntensif Aquitain (MCIA).
\end{acknowledgement}

\begin{suppinfo}
\end{suppinfo}


\bibliography{erA}


\clearpage


\newpage


\begin{table}
\caption{Values of the adsortion energy ($Q_A$), ZPE along the
$x$-, $y$-, $z$-axes, and
cartesian coordinates of the most favorable adsorption sites 
for the H/W(100), H/W(110), N/W(100), and N/W(110)
systems. The origin of the coordinate system is located on a W surface atom.}
\label{tab:zpeadsit}
\begin{tabular}{ l c c c c c c c }
 \hline
& & \multicolumn{3}{c}{ZPE (meV)} & \multicolumn{3}{ c }{adsorption site ({\AA})} \\
{system} & {$Q_A$ (eV)} & {$x$-} & {$y$-} & {$z$-} & {$X$} & {$Y$} & {$Z$} \\
\hline
{H/W(100)} & {3.07} & {55} & {33} & {67} & {1.585} & {0.0} & {1.2} \\
{H/W(110)} & {3.07} & {55} & {55} & {68} & {1.585} & {0.634} & {1.096} \\
{N/W(100)} & {7.37} & {11} & {11} & {29} & {1.5874} & {1.5874} & {0.65} \\
{N/W(110)} & {6.86} & {28} & {12} & {38} & {1.5874} & {0.0} & {1.155} \\
\hline
\end{tabular}
\end{table}


\begin{table}
\caption{Rebound altitude of the projectile for the BOSS model when E$_i=$1.0~eV (4.0~eV).}
\label{tab:rebz}
\begin{tabular}{ c c c c }
\hline
H$+$H/W(100) & H$+$H/W(110) & N$+$N/W(100) & N$+$N/W(110) \\
\hline
0.26~{\AA} (0.24~{\AA}) & 0.58~{\AA} (0.46~{\AA}) & 0.61~{\AA} (0.66~{\AA}) & 1.26~{\AA} (1.29~{\AA}) \\
\hline
\end{tabular}
\end{table}

\newpage



\newpage


\newpage


\newpage


\end{document}